\documentclass[%
 reprint,
 amsmath,amssymb,
 aps,
]{revtex4-2}

\usepackage{comment}
\usepackage[normalem]{ulem}
\usepackage{graphicx}
\usepackage{dcolumn}
\usepackage{bm}
\usepackage{hyperref}

\usepackage{soul,xcolor}
\setstcolor{red}

\begin{document}

\title{Eigenvalue crossing as a phase transition in relaxation dynamics}
\author{Gianluca Teza}%
\author{Ran Yaacoby}
\author{Oren Raz}

\email{oren.raz@weizmann.ac.il}
\affiliation{Department of Physics of Complex Systems, Weizmann Institute of Science, Rehovot 7610001, Israel}

\date{\today}

\begin{abstract}
When a system's parameter is abruptly changed, a relaxation towards the new equilibrium of the system follows. We show that a crossing between the second and third eigenvalues of the relaxation matrix results in a relaxation trajectory singularity, which is analogous to a first-order equilibrium phase transition. We demonstrate this in a minimal 4-state system and in the thermodynamic limit of the 1D Ising model.
\end{abstract}

\maketitle

Our understanding of out-of-equilibrium systems primarily developed through analogies with systems at thermal equilibrium \cite{bouchaud1990anomalous,Touchette2009}.
In equilibrium, phase transitions are associated with a non-analytic behavior of the free energy. The singular character of phase transitions contributed to a strong and rigorous consolidation of such parallels \cite{stella2022anomalous}, with universal features establishing profound connections among apparently far and unrelated phenomena. 
Much effort was invested throughout the last century in understanding first- and second-order equilibrium phase transitions. This study led to the advancement of several important results, including the exact solution of the 2D Ising model \cite{baxter2016exactly}, the phenomenological Ginzburg-Landau theory \cite{landau2013statistical}, the Yang-Lee zeros \cite{lee1952statistical} and even renormalization group \cite{binney1992theory,kardar2007statistical}. Many of these techniques were developed due to an inherent difficulty associated with phase transitions: short-range equilibrium systems do not have any phase transition unless the thermodynamic limit (where the number of particles goes to infinity) is taken, and even then, they do not exist if the system is ``too simple'', e.g., is one dimensional or consists of non-interacting particles \cite{pathria2016statistical}.    

Although phase transitions were originally studied in equilibrium systems, similar phenomena also appear away from thermal equilibrium.
In fact, most equilibrium phase transitions have a dynamical counterpart when the external parameter changes across or is quenched through its critical value \cite{zamora2020kibble,Holtzman2022}. {Several nonequilibrium effects have similar characteristics to equilibrium phase transitions.
For example, the same power law singularities of second-order phase transitions observed in many equilibrium systems at criticality can be found in generating functions of diffusing systems \cite{stella2022anomalous,horowicz2021critical}.
Dynamical phase transitions \cite{vaikuntanathan2014dynamic,nemoto2017finite,nyawo2017minimal,nyawo2018dynamical} and discontinuities in the large-deviation rate function \cite{baek2018dynamical,baek2015singularities,meibohm2022finite,buvca2019exact} also withstand similar analogies.} In contrast to the equilibrium case, a nonequilibrium phase transition might have a constant flux across it \cite{nakagawa2017liquid}. These dynamical phase transitions often have different characteristics than their equilibrium counterpart. In addition, non-equilibrium systems can have phase transitions even under conditions that are incompatible with phase transitions at equilibrium, for example in 1D systems \cite{miron2020phase,majumdar1998nonequilibrium,vladimir1997nonequilibrium}.

In this letter, we show how \emph{eigenvalue crossing} between the second and third eigenvalues of a Markovian operator can induce a singularity in the long time limit approach to equilibrium as a function of the bath temperature $T_b$. As in dynamical phase transitions \cite{vaikuntanathan2014dynamic,nemoto2017finite,nyawo2017minimal,nyawo2018dynamical}, here the long-time limit $t\to \infty$ replaces the thermodynamic limit $N\to\infty$, but unlike dynamical phase transitions, the discontinuity is not in the probability of observing a rare-event, but rather in the average direction of the relaxation to equilibrium. Similar to level crossing in quantum systems \cite{landau2013quantum}, this phase transition requires some symmetry in the system, otherwise, small perturbations make the exact crossing turn into \emph{avoided crossing}. However, even avoided crossing is a sufficient condition to explain the appearance of other anomalous relaxation phenomena, like the Mpemba effect (ME) \cite{Mpemba1969,Lu2017,Klich2019,Gal2020,Lasanta2017,busiello2021inducing,Walker2021,Kumar2020,kumar2022anomalous}.
We demonstrate our results in two systems: first, in the simplest system that can exhibit exact eigenvalue crossing -- a four-state system with Arrhenius rates, but where every perturbation results in avoided crossing. Then we consider the 1D antiferromagnet Ising chain, where the thermodynamic limit can be taken analytically, showing that the effect exists even in macroscopic many-body systems. Moreover, the two symmetries of the antiferromagnet Ising model protect the crossing against small perturbations.

It is instructive to start by considering why there are no equilibrium phase transitions in finite systems. One way to argue this is to use a detailed-balance Markovian rate matrix $R$ whose steady-state distribution is the {Boltzmann} equilibrium distribution, e.g., using Glauber rates for the dynamics \cite{Glauber1963}. In this case, the equilibrium distribution is the eigenvector corresponding to the largest eigenvalue of the matrix $e^{Rt}$ for any $t$.
{The Perron-Frobenius theorem \cite{ninio1976simple} ensures that the largest eigenvalue of $e^{Rt}$ is non-degenerate, and therefore there cannot be an eigenvalue crossing for any value of the parameters.} This implies that for rates that are analytic in the external parameters, the corresponding eigenvector is also an analytic function of these parameters, and there is no phase transition in the system.
However, the same argument does not hold for the case of an infinite system since the Perron-Frobenius theorem cannot be applied for the corresponding Markovian operator \cite{ninio1976simple}, and hence a phase transition is possible in the thermodynamic limit.
A key point in our {analysis} is that for relaxation processes, it is not the largest eigenvalue of the Markovian matrix that controls the process (it only controls the final state), but rather the rest of them.
Specifically, the long-time limit of the relaxation process is controlled by the second eigenvalue, and as discussed below, a crossing between the second and third eigenvalues generates a singularity in the relaxation dynamics.

To observe eigenvalue crossing, we have to track how they change as a function of some parameter. For concreteness, we choose to use here the external bath temperature $T_b$, but a similar analysis can be performed for any other external parameter. Limiting our discussion to discrete setups for simplicity, the system is described by a vector $p_i$, indicating the probability of observing the system in a certain microscopic configuration $i$. The evolution of the system is stochastic, and the probability distribution $p_i$ evolves by the master equation
\begin{equation} \label{eq:rate_eq}
    \partial_t\vec{p}(t) = R (T_b)\vec{p}(t),
\end{equation}
where $R (T_b)$ is the rate matrix containing all the specific details of the system and its coupling to the bath. The off-diagonal elements $R_{ij}$ are the jump rates from microstate $j$ to $i$, while $R_{ii} = - \sum_{j \neq i} R_{ji}$ represent the escape rates from the state $i$.
Assuming that $R$ is irreducible and satisfies detailed balance, the system eventually relaxes towards the (unique) Boltzmann equilibrium $\pi_i (T_b) = e^{- E_i/T_b}/Z(T_b)$, where $E_i$ is the energy of the microstate $i$, and $Z(T_b) = \sum_i e^{- E_i/T_b}$ is the partition function (we use units in which $k_B=1$).
Formally integrating Eq. \ref{eq:rate_eq} with a Boltzmann equilibrium at temperature $T_0$ as initial condition gives
\begin{equation} \label{eq:prob}
    \vec{p}(t, T_b, T_0) = \vec{\pi}(T_b) + \sum_{n>1} a_n (T_b, T_0) e^{\lambda_n (T_b) t} \vec{v}_n (T_b)\ ,
\end{equation}
where $\vec{v}_n$ are the right eigenvectors of $R$ with associated real \footnote{The eigenvalues are all real-valued since $R$ satisfies the detailed balance conditions.} eigenvalues $0=\lambda_1> \lambda_2 \ge \lambda_3 \ge \dots$ and the coefficients $a_n$ correspond to the projections of the initial state on the left eigenvectors. While $\lambda_1=0$ is granted to be non-degenerate in such systems, the same does not apply to all the remaining eigenvalues.

\begin{figure}
    \includegraphics[width=\linewidth]{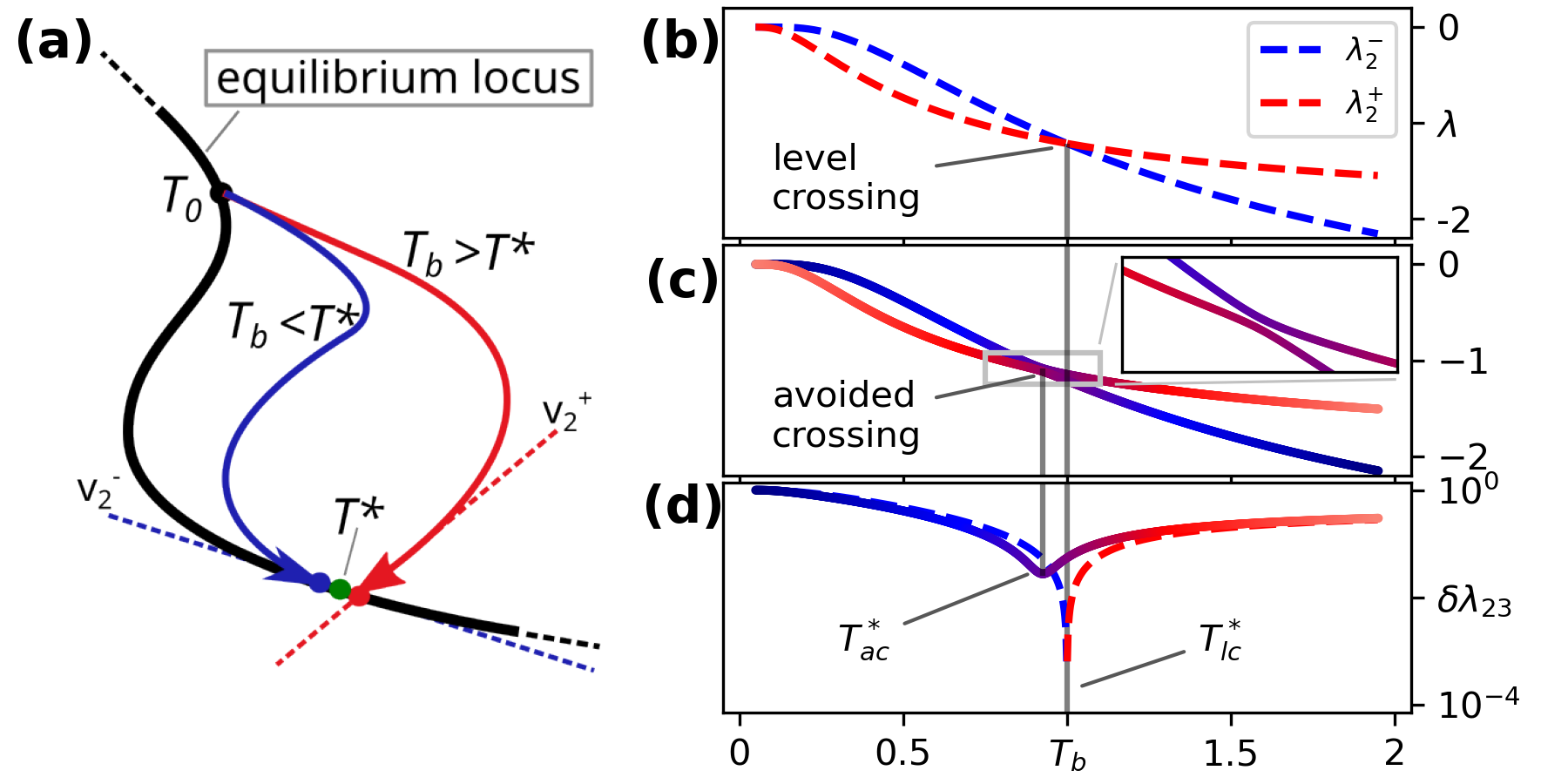}
    \caption{(a) A small change in the bath temperature $T_b$ drastically changes the approach to equilibrium direction in the presence of a level crossing (b) at a certain temperature $T^*$, like the one observed in the four-state system presented in the text (Eq. \ref{eq:arrhenius_form}).
    (c) A non-singular transition occurs in the case of avoided crossing, where the relative eigenvalue difference $\delta\lambda_{23}$ exhibits a minimum rather than a marked dip at the crossing temperature (d).}
    \label{fig:Fig1}
\end{figure}

The second eigenvalue represents the slowest dynamics, setting an exponential timescale of the relaxation $\propto1/|\lambda_2|$.
Indeed, substituting the formal solution in the master equation gives
\begin{equation}\label{eq:app_dir_p}
    e^{-\lambda_2t}\partial_t\vec{p}(t) = a_2\lambda_2\vec{v}_2 + \sum_{n>2}a_n\lambda_n e^{-\Delta\lambda_{2,n} t}\vec{v}_n 
\end{equation}
where we introduced the eigenvalue gaps $\Delta\lambda_{2,n} = \lambda_2-\lambda_n\geq0$.
If $\lambda_2$ is not degenerate and $a_2\neq 0$, the final stage of the relaxation is in the direction of $\vec{v}_2$, and it changes continuously with $T_b$. However, $\vec{v}_2(T_b)$ can abruptly change at some temperature $T^*$ if at such temperature there is an eigenvalue crossing, namely $\lambda_2(T^*) = \lambda_3(T^*)$ as in Fig. (\ref{fig:Fig1}). This eigenvalue crossing is algebraically identical to  \emph{level crossing} in the context of diabatic passages of a two-level quantum system \cite{Landau1981,Kato1950}, though its implications are different: in the $t\to\infty$ limit, the crossing implies a jump in the final direction of the approach to equilibrium; thus it can be interpreted as a phase transition in the relaxation dynamics. 

Referring to the roots and eigenvectors that dominate the long-time dynamics before and after $T^*$ as $\lambda_2^{\pm}$ and $\vec{v}_2^{\pm}$ (Fig. \ref{fig:Fig1}a), we characterize the singular behavior in the long-time limit of Eq. \ref{eq:app_dir_p} as
\begin{equation}
\vec{v}_2 = 
    \begin{cases}
    \vec{v}^{-}_2 & T_b<T^* \\
    a_2^-\vec{v}^{-}_2 +  a_2^+\vec{v}^{+}_2 & T_b=T^* \\
    \vec{v}^{+}_2 & T_b>T^* \\
    \end{cases}
\end{equation}
where $a_2^{\pm}$ are coefficients {determined} by the initial conditions.
Note that this singularity is not detectable in the equilibrium steady state but rather in the relaxation to equilibrium dynamics of the system. This is why eigenvalue crossing can be linked with anomalous phenomena arising in the relaxation process \cite{Walker2021,Teza2021_mpemba_boundary}.

We note that in the original solution to the 2D Ising model, due to Onsager, the phase transition temperature was also found through the point at which the largest eigenvalue of the transfer matrix becomes degenerate \cite{baxter2016exactly}.
However, this degeneracy is not an eigenvalue crossing: the largest eigenvalue continues to be degenerate for all temperatures below the critical one, and the two degenerate eigenvectors correspond to the two phases, as expected in a second-order phase transition.
Therefore, we interpret the eigenvalue crossing in the relaxation dynamics as a first-order phase transition. It is possible to have a second-order phase transition in systems with broken detailed balance, when the rate matrix passes through an exceptional point beyond which $\lambda_2$ becomes complex-valued and hence $\textrm{Re} (\lambda_2) = \textrm{Re} (\lambda_3)$.

Algebraically, eigenvalue crossing is not stable since the dimension of matrices with level crossing is smaller than the dimension of all relevant matrices. Thus, unless some symmetry prevents perturbations in the rate matrix in the direction that {breaks} the degeneracy, the singular phase transition is not expected to be directly observed.
However, even in this case, the non-degeneracy of the second eigenvalue induces a sharp -- albeit non-singular -- transition across $T^*$ in the approach to equilibrium direction (see Fig. \ref{fig:Fig1}c).
The timescale of the slowest dynamics is then regulated by the relative difference $\delta\lambda_{23}=-\Delta\lambda_{23}/\lambda_2>0$, as it can be easily seen by rescaling the time by $\lambda_2$ in Eq. \ref{eq:app_dir_p}.

\begin{figure}
    \includegraphics[width=\linewidth]{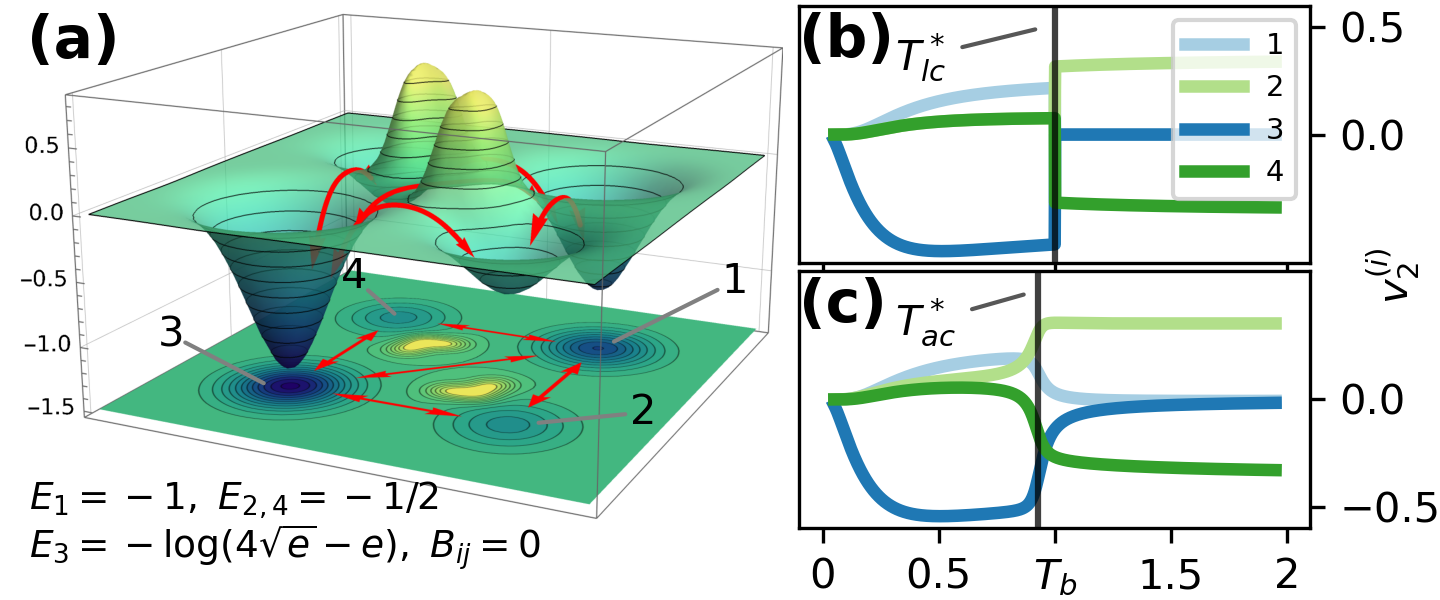}
    \caption{(a) Four-wells energy landscape associated with the level crossing (Fig. \ref{fig:Fig1}b) Markov model presented in the text.
    Red arrows indicate the allowed transitions among the wells.
    (b) The system exhibits a second order transition in the dynamics at $T^*_{lc}=1$.
    The components of $\vec{v}_2$ are numbered according to the 4 wells.
    (c) Breaking the energy degeneracy by setting $E_2=-0.6$ smoothens the transition, providing the avoided crossing of Fig. \ref{fig:Fig1}c around $T^*_{ac}\sim0.93$.}
    \label{fig:Fig2}
\end{figure}

Let us present the minimal model that exhibits eigenvalue crossing in the dynamics: a four-state Markovian system coarse-grained from an overdamped four-well energy landscape (Fig. \ref{fig:Fig2}a).
Indeed, it can be proven that any $N$-state system with an $N-1$ degeneracy of $\lambda_2$ at a certain temperature $T^*$ necessarily extends to a degeneracy for every bath temperature \cite{SM}, ruling out level crossing in three-state systems.
A general representation of the rate matrix $R$ is the Arrhenius form \cite{Mandal2011,raz2016mimicking}:
\begin{equation} \label{eq:arrhenius_form}
    R_{ij}(T_b) = 
    \begin{cases}
        \Gamma e^{-(B_{ij}-E_j)/T_b} & i\neq j\\
        -\sum_{k \neq i} R_{ki} & i = j
    \end{cases}
\end{equation}
where $\Gamma$ ($=1$ for simplicity) is a rate constant and $B_{ij} = B_{ji}$ denotes the energy barrier between state $i$ and $j$, set to be higher than the adjacent energy levels $E_{\{i,j\}}$. Given the low number of free parameters in a four-state system, one can easily find an example with eigenvalue crossing, as explained in the SI \cite{SM}. {In our specific example, all transitions are permitted through finite barriers at height $B_{ij}=0$, apart from the one between two opposite degenerate wells ($E_2=E_4=-1/2$). In addition we set $E_1=-1$, while $E_3=-\log(4\sqrt{e}-e)$ is determined by the constraints of a crossing at $T^*=1$.}
This example exhibits a marked crossing (Fig. \ref{fig:Fig1}b) that induces a phase transition in the relaxation dynamics at $T^*$, as we can see through the components of $\vec{v}_2$ depicted in Fig. \ref{fig:Fig2}b. The singularity can also be characterized by the relative eigenvalue difference $\delta \lambda_{23}$, which exhibits a marked dip at the crossing temperature (Fig. \ref{fig:Fig1}d).

A minor perturbation in the parameter values generally converts the singularity into avoided crossing. For instance, breaking the energy degeneracy by setting $E_2=-0.6$ results in the avoided crossing shown in Fig. \ref{fig:Fig1}c, which nevertheless induces a sharp but continuous transition of $\vec{v}_2$ (Fig. \ref{fig:Fig2}c) provided that the dimensionless timescale $\delta \lambda_{23} \ll 1$ (Fig. \ref{fig:Fig1}d). This feature is fundamental when considering experimental setups in which one might want to detect this phenomenon. Indeed, the parameters can be tuned only to within a certain precision depending on specific details of the experimental apparatus. This result not only increases the chances of observing the effect considerably but also opens up the possibility of seeing it in even simpler setups as a three-state system; see \cite{SM}.

The sensitivity of the eigenvalue crossing to small perturbations is not important in highly symmetric models if the symmetry prevents perturbations that break the degeneracy. An example of such a system is the 1D Ising antiferromagnet chain.
Consider a ring of $N$ spins $\sigma_s=\pm1$, for which the Hamiltonian for any configuration $\vec{\sigma}$ of the $2^N$ possible microstates reads
\begin{equation}\label{eq:ham_ising}
    \mathcal{H}(\vec{\sigma})=-H\sum_{s=1}^N \sigma_s-J\sum_{s=1}^N \sigma_s \sigma_{s+1}
\end{equation}
where $H$ is the magnetic field, $J<0$ is the antiferromagnetic coupling constant and $\sigma_{N+1}\equiv\sigma_1$.
We implement single-spin Glauber dynamics \cite{Glauber1963}, namely the rates connecting two microscopic configurations $\vec{\sigma}^{\{i,j\}}$ with energies $E_{\{i,j\}}$ is
\begin{equation}\label{eq:glauber}
    R_{ij}=\frac{\delta_{1,d_{ij}}}{1+e^{(E_i - E_j)/T_b}}
\end{equation}
where $d_{ij}=\sum_s \delta_{\sigma_{s}^i,-\sigma_{s}^j}$ and the Kronecker delta function $\delta_{1,d_{ij}}$ limits the transition to single-spin flips.

\begin{figure}
    \centering
    \includegraphics[width=\linewidth]{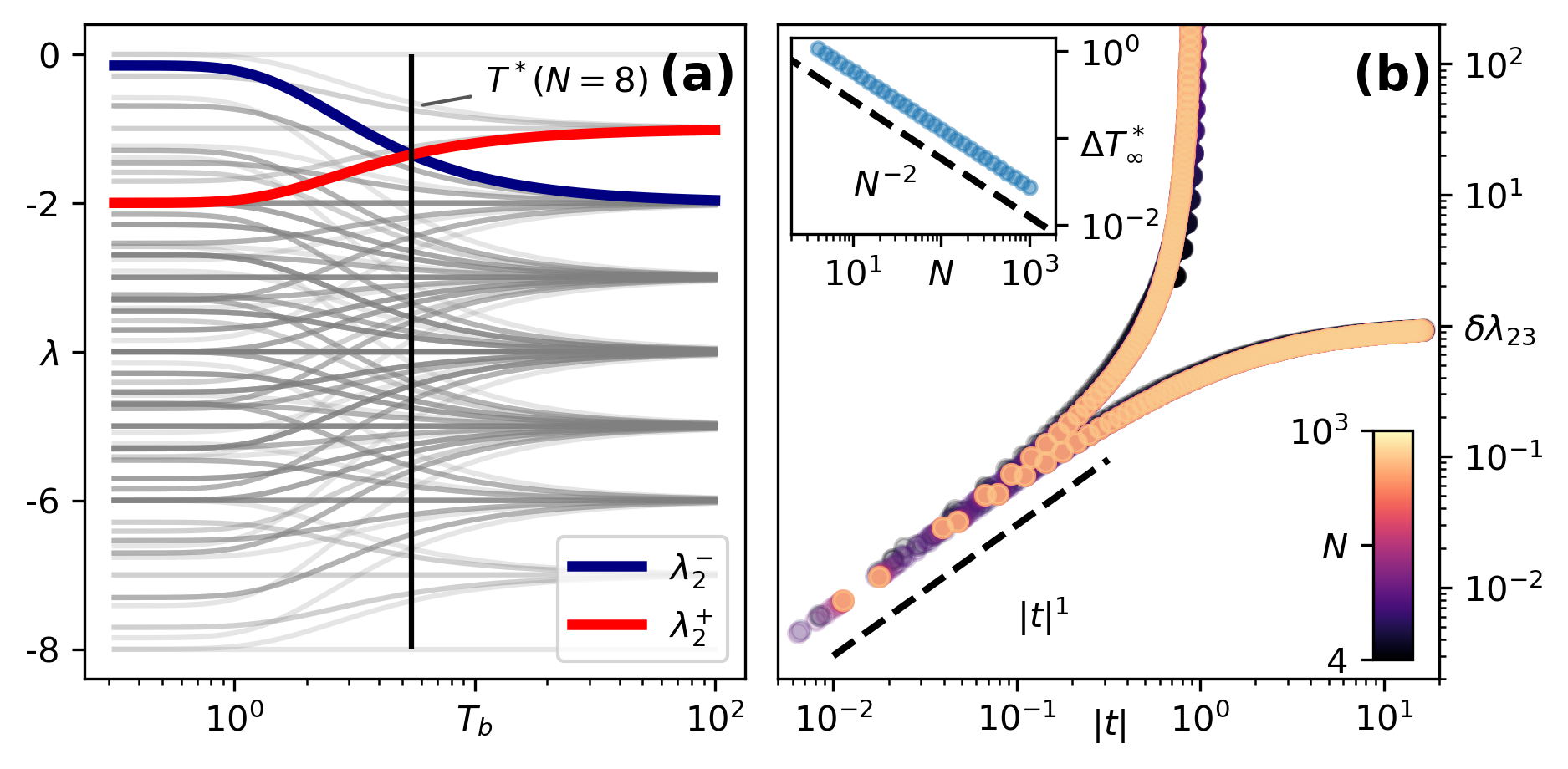}
    \caption{(a) Spectrum of an $N=8$ Ising antiferromagnet at zero magnetic field with Glauber dynamics (Eq. \ref{eq:glauber}).
    Highlighted in blue and red are the first two dominant eigenvalues which are relevant for the dynamics. They cross at  $T^*(N=8)\sim5.45$.
    (b) Eigenvalue difference $\delta\lambda_{23}$ as a function of the {reduced} temperature $t=(T-T^*)/T*$, exhibiting an excellent finite-size collapse $\propto|t|^1$ around the crossing temperature.
    Inset: the distance from the asymptotic crossing temperature $T^*_\infty=2/\mathrm{arctanh }(1/3)$ decays quadratically.
    }
    \label{fig:Fig3}
\end{figure}
In the absence of magnetic field $H=0$, an explicit form for all the eigenvalues and eigenvectors was derived by Felderhof \cite{Felderhof1971}. Introducing the sets $S^{+}=\{\pi \frac{2i-1-N}{N}\}$ and $S^{-}=\{\pi \frac{2i-N}{N}\}$ with $i=1\dots N$, and considering all the $k$-combinations $\vec{q}_k\in\binom{S^{\pm}}{k}$ where $S^{+}$ ($S^{-}$) is chosen for even (odd) values of $k$, we can express the $2^N -1$ eigenvalues regulating the dynamics as
\begin{equation}
    \lambda(\vec{q}_k)=-k+\gamma \sum_{i=1}^k\cos (q_{k,i})
\end{equation}
 where $\gamma=\tanh 2J/T_b$.
This system is invariant with respect to symmetries that considerably reduce the number of eigenvalues relevant to the dynamics. The set of all rate matrices, under global flipping and cyclic shifts of the microscopic configurations is isomorphic to the $Z_2\times D_N$ group  \footnote{With $Z_2$ we refer to the cyclic group of order 2, while $D_N$ is the dihedral group of the symmetries of a regular $N$ sided polygon.}.
Therefore, in the antiferromagnetic case ($J<0$), we find that the first eigenvalues of eigenvectors with even parity with respect to such symmetry turn out to be $\lambda_2^-=-2+2\gamma \cos \left( \frac{N-1}{N} \pi \right)$ and $\lambda_2^+=-1+\gamma$, highlighted in Fig. \ref{fig:Fig3}a.
The analytic expressions of the eigenvalues enable us to formally study the phenomenon in the thermodynamic limit: imposing $\lambda_2^-=\lambda_2^+$, we find that the crossing survives the $N\to\infty$ limit, asymptotically approaching $T^*_{\infty}=2/\mathrm{arctanh }(1/3)$ (Fig. \ref{fig:Fig3}b). The eigenvalue difference exhibits also an excellent finite-size multi-scale collapse \cite{Teza2019} against the {reduced} temperature $t=(T-T^*)/T^*$ with a dependence $\propto|t|^1$, while the distance from the asymptotic crossing temperature $T^*_{\infty}$ decays quadratically.

\begin{figure}
    \centering
    \includegraphics[width=\linewidth]{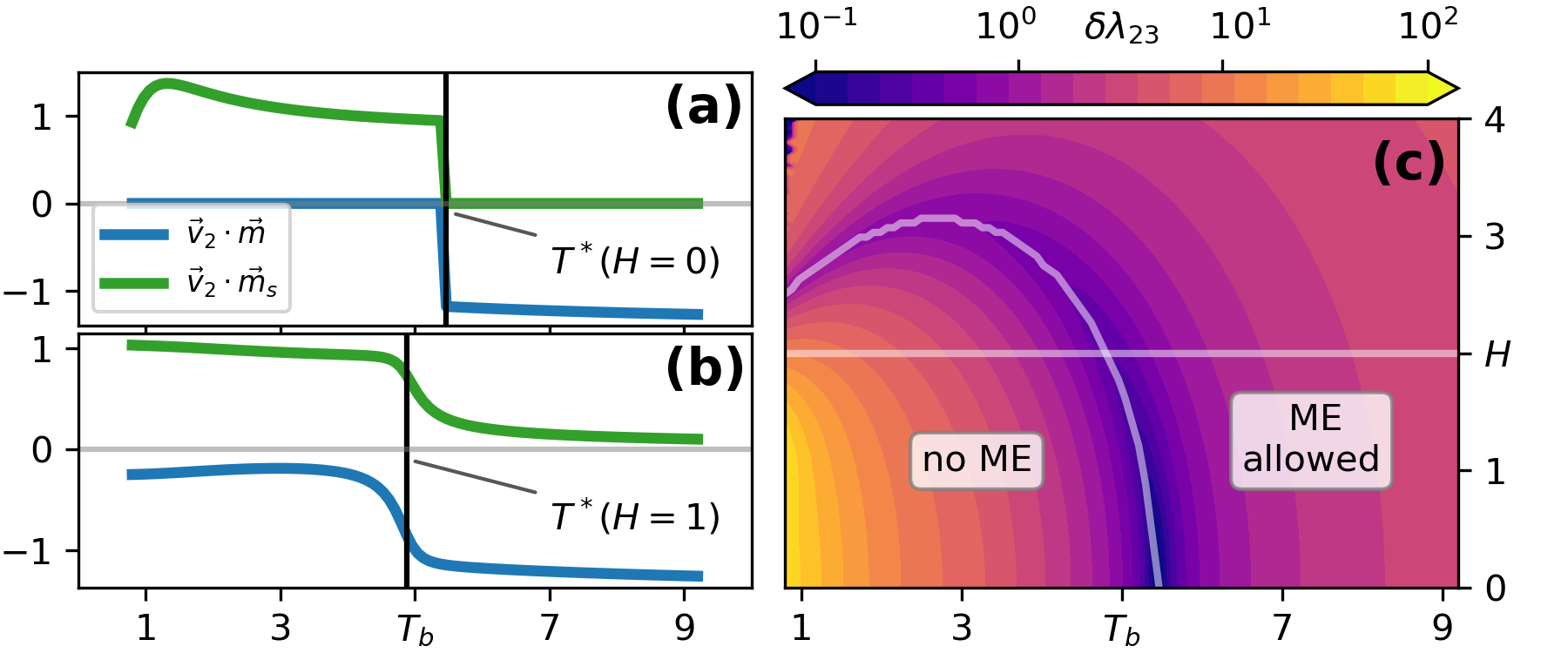}
    \caption{(a) Projection of $\vec{v}_2$ onto the magnetization $\vec{m}$ and staggered magnetization $\vec{m}_s$ vectors. At $H=0$ the relaxation is completely orthogonal to $\vec{m}$ ($\vec{m}_s$) for all temperature below (above) $T^*$.
    (b) At $H>0$ the transition becomes non-singular.
    (c) The eigenvalue difference $\delta\lambda_{23}$ highlights the avoided crossing (white curved line) induced on the system through a magnetic field perturbation.
    The horizontal white line delimits the ferromagnetic phase ($|H|<2$), where the ME is found to be allowed in the right region, in which $\vec{v}_2\parallel\vec{m}$.}
    \label{fig:Fig4}
\end{figure}

The eigenvector directions associated with the crossing eigenvalues have a {clear}  physical meaning in this system. The $2^N$-dimensional eigenvectors can be projected along the magnetization and staggered magnetization vectors, defined as $(\vec{m})_i=\sum_s \sigma^i_s$ and $(\vec{m}_s)_i=|\sum_s (-1)^s \sigma^i_s|$ for a given microscopic configuration $\vec{\sigma}^i$ \cite{Kampen2007}. In Fig. \ref{fig:Fig4}a we show the projection of $\vec{v}_2$ along such directions, finding that it is identically zero before (after) the crossing temperature along $\vec{m}$ ($\vec{m}_s$). This indicates that the approach to equilibrium occurs along the staggered magnetization for bath temperatures $T_b<T^*$ and along the magnetization for $T_b>T^*$, while at $T_b=T^*$ it follows along some linear combination of $\vec{m}$ and $\vec{m}_s$ depending on the initial conditions. Any perturbation that does not break the two symmetries associated with these eigenvectors -- flipping all the spins or translating the chain by a single spin position -- would not split the eigenvalue crossing.
However,  perturbing the system, for example, with a magnetic field $H>0$ breaks the singular behavior smoothing the transition, which can nevertheless be arbitrarily sharp for small enough magnetic fields (Fig. \ref{fig:Fig4}b).

The eigenvalue crossing is one of the possible mechanisms by which the Mpemba parity index \cite{Klich2019}, which is a topologically protected quantity, can nevertheless change. Indeed, in many variants of the anti-ferromagnetic Ising model at $H=0$ there is a sharp transition at some temperature from zero to non-zero Mpemba index  \cite{Klich2019,Teza2021_mpemba_boundary,Teza2022,Holtzman2022}.
This was already pointed out in Ref. \cite{Teza2021_mpemba_boundary}, where an exact coarse-graining procedure \cite{Teza2020,Teza2020b} allowed to explore large-sized systems and to argue that the effect survives in the thermodynamic limit.
In Fig. \ref{fig:Fig4}c we plot the eigenvalue difference $\delta\lambda_{23}$ as a function of both bath temperature and magnetic field in the 1D antiferromagnetic Ising model. The antiferromagnetic phase for $J=-1$ is delimited by $|H|<2$ (horizontal white line), corresponding to the region in which the external magnetic field is not strong enough to overcome the negative nearest neighbor interaction among the spins. The minima of $\delta\lambda_{23}$ (curved white line) partition the parameters space, showing that the existence of the ME is limited to the region in which $\vec{v}_2\parallel\vec{m}$. Anomalous relaxation effects can also be observed for $|H|>2$. Still, their appearance in the 1D system is related to finite size effects and is therefore not expected to survive the thermodynamic limit \cite{Teza2021_mpemba_boundary}.

Summarizing, we have shown how eigenvalue crossing can be interpreted as a phase transition in the dynamics of stochastic systems. Such a transition can drastically change the direction from which the system approaches the bath temperature equilibrium, thereby explaining where anomalous relaxation effects can be observed in terms of model parameters. It was shown that eigenvalue crossing appears in the paradigmatic 1D Ising antiferromagnet, and it survives in the thermodynamic limit, with relaxation occurring along the staggered (total) magnetization before (after) the crossing. We have shown how an external perturbation breaks the singularity in the dynamics but nevertheless maintains a steep jump related to a marked avoided crossing. This is important when attempting to observe this phenomenon in simpler, single-body experimental setups, where model parameters can be tuned only up to some finite precision. The four-state example we provided not only serves as a pedagogical example but also provides the means to characterize this phenomenon in small experimental setups, such as the colloidal systems in which the ME was recently observed \cite{Kumar2020,kumar2022anomalous}.

\begin{acknowledgments}
O. R. is the incumbent of the Shlomo and Michla Tomarin career development chair and is supported by the Abramson Family Center for Young Scientists, the Israel Science Foundation Grant No. 950/19 and by the Minerva foundation.
G. T. is supported by the Center for Statistical Mechanics at the Weizmann Institute of Science, the grant 662962 of the Simons foundation, the grants HALT and Hydrotronics of the EU Horizon 2020 program and the NSF-BSF grant 2020765.
We thank David Mukamel, John Bechhoefer and Attilio L. Stella for useful discussions.
\end{acknowledgments}

\bibliography{refs.bib}

\end{document}